\documentclass[12pt,nofootinbib]{revtex4}
\usepackage{times}
\usepackage{hyperref}
\usepackage[justification=centering]{caption}
\usepackage[utf8]{inputenc}
\usepackage{amssymb,amsmath} 
\usepackage[T1]{fontenc}
\usepackage{graphicx}
\usepackage{color}
\usepackage{t-angles}
\usepackage{braket}
\usepackage{mleftright}
\usepackage{diagbox}
\begin{document} 
\title{Finite pseudo-Riemannian spectral triples and the standard model.} 
\author{Arkadiusz Bochniak${}^\dagger$}
\author{Andrzej Sitarz${}^\dagger$${}^\ddagger$}
\thanks{Authors acknowledge support by NCN grant OPUS 2016/21/B/ST1/02438}
\affiliation{${}^\dagger$Institute of Physics, Jagiellonian University,
	\hbox{prof.\ Stanis\l awa \L ojasiewicza 11, 30-348 Krak\'ow, Poland} \\ ${}^\ddagger$Institute of Mathematics of the Polish Academy of Sciences,
	\'Sniadeckich 8, 00-656 Warszawa, Poland.}
\begin{abstract} 
We interpret the physical symmetry preserving the lepton number as a 
shadow of a finite pseudo-Riemannian structure of the standard model.
Using the pseudo-Riemannian generalizations of real spectral triples we 
describe the geometries with indefinite metric over finite-dimensional 
algebras and their Riemannian shadows. We apply the discussion to
the standard model spectral triple, and classify possible time orientations
leading to restrictions on the physical parameters and symmetries.
\end{abstract} 
\maketitle 
\section{Introduction}
Noncommutative geometry offers an intriguing possibility of a new insight into the structure of all fundamental interactions,
linking purely geometric gravity with the electroweak and strong interactions of elementary fermions \cite{con}. The possibility
is explored in one way into extending the geometric notions to describe models that could approximate spacetime \cite{madore} and,
on the other hand, to gather the information about the structure of geometry underlying experimentally verified models 
of fundamental constituents of matter \cite{grav, wvs}.

The crucial role in the understanding of the geometric interpretation of the standard model of particle physics is based on
the finite geometry, linked to the finite-dimensional algebra $\mathbb{C} \oplus  \mathbb{H} \oplus M_3(\mathbb{C})$ 
and the related finite spectral triple. Even though finite spectral triples have been classified some time ago \cite{ps, kraj} and
the model has been extensively studied, it still can surprise and shed new light on the structure of fundamental interaction. 
An example is the recent discovery of unexpected duality \cite{das} in the standard model Clifford algebra (called Hodge duality) that is satisfied only for certain values of physical parameters (bare masses and mixing matrices). 

Though the quest for the better understanding of the structure has already brought new results, some issues still remain 
unsolved, like the consistent Lorentzian framework for standard model description \cite{ps-lor, barret, franco}, the fermion doubling problem \cite{lizzi1, doubl} or the classification of possible Dirac operators \cite{boyle, bizi}. The latter appears to be, so far, the most important issue, as even with the requirement of some additional symmetries (second-order condition \cite{das} later incorporated as originating from the Hodge duality \cite{d}) there exist Dirac operators that allow for the $SU(3)$ symmetry breaking and lead to the unphysical leptoquarks \cite{PSS}. 

In this paper we propose an alternative explanation of the observed quarks-leptons symmetry which prevents the $SU(3)$-breaking,
as a shadow of the Lorentzian structure. We propose also that the consistent model-building for the physical interactions and 
possible extensions of the standard model within the noncommutative geometry framework should use possibly the pseudo-Riemannian
extension of finite spectral triples, for which we present a consistent and clear framework. We demonstrate that the pseudo-Riemannian
framework allows for more restrictions and, in the discussed case introduces an extra symmetry grading, which we interpret as the lepton-quark symmetry (that was postulated as the so-called $S_0$-symmetry in \cite{c, co, kraj}). The physical
interpretation of this symmetry is the lepton number conservation, which is
strongly confirmed by current experimental data \cite{pd}.

To finish the introduction let us briefly describe the notation and mathematical
constructions used in this paper. We consistently use particle physics 
convention with positive sign of the metric for the time direction and 
negative sign for spatial directions. We use the notion of a Clifford 
algebra, which is a matrix algebra that encodes the $\gamma$ matrices 
of the Dirac operator. The definition of the Clifford algebra is taken so that for 
a vector space with a quadratic form of signature $(p,q)$ it is generated by 
$p$ matrices of square $1$ and $q$ of square $-1$ that anticommute with 
each other. We work with a complexified Clifford algebra, which represented on
a space of complex spinors, however, with the real structure of the Clifford
algebra encoded through an antilinear operator on the space of spinors. 
The signature of the metric is visible in an additional structure on the spinor
space, which gives rise to an indefinite scalar product (called Krein product). 
Details of these constructions and definitions are given in Sec. \ref{cliff}.

\section{Pseudo-Riemannian spectral triples}

Let us recall that a real pseudo-Riemannian spectral triple of signature $(p,q)$ is a system $(\mathcal{A},\pi,\mathcal{H},D,J,\gamma,\beta)$ where 
$\mathcal{A}$ is an involutive unital algebra, $\pi$ its faithful $\ast$-representation on 
an Hilbert space $\mathcal{H}$ such that the following conditions hold.
First, for even $p\!+\! q$ there exists a  $\mathbb{Z}_2$-grading $\gamma^\dagger=\gamma,\gamma^2=1$ commuting with the representation of 
$\mathcal{A}$, $J$ is an antilinear isometry and for all $a,b \in \mathcal{A}$ we have $[J \pi(a^*) J^{-1},  \pi(b)]=0$. Furthermore, there
exists an additional grading $\beta= \beta^\dagger, \beta^2=1$ also commuting with the representation of $\mathcal{A}$, which defines the 
Krein structure on the Hilbert space. The latter is an indefinite bilinear form defined
as $(\phi,\psi)_\beta = (\phi, \beta \psi)$, where $(\cdot,\cdot)$ is the usual positive
definite scalar product on the Hilbert space. As a last requirement, we postulate 
the existence of a (possibly unbounded) densely defined operator $D$, 
which is $\beta$-self-adjoint, i.e. $D^\dagger=(-1)^p\beta D\beta$ and such that 
$[D,\pi(a)]$ is bounded for every $a\in\mathcal{A}$, is odd with
respect to $\gamma$-grading: $D\gamma=-\gamma D$. The operators 
$D,\gamma,J$ satisfy following (anti)commutation relations, which
depend on the signature of the pseudo-Riemannian space through $p\!-\!q$ modulo $8$:
\begin{equation}
DJ=\epsilon JD,\  J^2=\epsilon' \mathrm{id}, \ J\gamma=\epsilon'' \gamma J,
\end{equation}
where $\epsilon,\epsilon',\epsilon''=\pm 1$ are given in the Table \ref{table1}:
\begin{table}[h!tb]
	\caption{Signs for $KO$-dimension (mod $8$).}
	\label{table1}
  \begin{tabular}{ | c | c | c | c | c | c | c | c | c | }
    \hline
    $p\!-\!q \ \mathrm{mod} \ 8$ & $\; 0 \;$ & $\; 1\;$&$\; 2\; $&$\; 3\; $&$\; 4\; $&$\; 5\; $&$\; 6\; $&$\; 7\; $\\ \hline
    $\epsilon$ &$+$&$-$&$+$&$+$&$+$&$-$&$+$&$+$\\ \hline
    $\epsilon'$ &$+$&$+$&$-$&$-$&$-$&$-$&$+$&$+$ \\ \hline
    $\epsilon''$ &$+$&&$-$&&$+$&&$-$& \\ \hline 
\end{tabular}

\end{table}

The number $p\!-\!q$ modulo $8$, which determines the signs $\epsilon, \epsilon', \epsilon''$,  is called $KO$-dimension of the spectral triple 
as it relates to the real $K$-theory and periodicity (modulo $8$) in real Clifford algebras.

The Krein structure $\beta$ satisfies alone relations which depend only on $p$: 
\begin{equation}\beta\gamma=(-1)^{p}\gamma\beta, \ \beta J =(-1)^{\frac{p(p-1)}{2}} \epsilon^p J\beta. 
\end{equation}

Finally, to implement the condition that for manifolds $D$ is a first order differential 
operator, we impose the $1$st-order condition, requiring that for all 
$a,b \in \mathcal{A}$ 
\begin{equation}
\left[J\pi(a)J^{-1},[D,\pi(b)]\right]=0.
\end{equation}

A spectral triple defined above is orientable if there exists a finite collection of
elements from the algebra (which could be together combined in a so-called
Hochschild cycle of dimension $n\!=\!p\!+\!q$),  
$( a^i,  a_0^i,  a_{1}^i, \ldots, a_{n}^i), i=1,\ldots,k$ such that
\begin{equation}
\sum_{i=1}^k \left( J\pi\left(a^i \right)J^{-1} \right) \pi\left(a_{0}^i\right)\left[D,\pi\left(a_{1}^i\right)\right]...\left[D,\pi\left(a_{n}^i\right)\right]=
\begin{cases} \gamma & n \; \hbox{even}, \\ 1 & n \; \hbox{odd}. \end{cases}
\end{equation}
The orientation corresponds to the existence of the nowhere vanishing 
volume form.

In the pseudo-Riemannian case we can also define a separate notion of time-orientation (even if there are multidimensional times, that is $p>1$) in
the following way. We say that $\beta$ is a time orientation if there exists 
a collection of elements from the algebra $( b^i,  b_0^i,  b_{1}^i, \ldots, b_{p}^i), i=1,\ldots,k$ such that
\begin{equation}
\beta =\sum_{i=1}^k \left(J \pi(b^i)J^{-1} \right) \pi(b_0^i)[D,\pi(b_1^i)] \cdots [D,\pi(b_p^i)].
\end{equation}
In the case of Lorentzian manifolds ($p=1$) the above notion of time orientability 
is equivalent to the existence of global timelike vector field.

Using $\beta$ we can define the operator $\braket{D}=\sqrt{\frac{1}{2}(DD^\dagger + D^\dagger D)}$, which is self-adjoint 
on $\mathcal{H}$. Then, if we require that $\braket{D}$ has compact resolvent and $[\braket{D},[D,\pi(a)]]$ is bounded 
for all $a\in\mathcal{A}$ we have a pseudo-Riemannian version of the regularity and spectral condition for the Dirac operator.
	
\subsection{Clifford algebras of arbitrary signature\label{cliff}}	

Before we proceed with the special case of finite-dimen\-sional triples, let us briefly recall the 
conventions and basic properties of Clifford algebras for a metric of indefinite signature $(p,q)$, 
which motivate the above definition. We include this short paragraph so that the note is complete 
and self-contained.

Let us take the algebra generated by $\gamma_a$, $a=1,\ldots,p\!+\!q$, with
the relations $ \gamma_a \gamma_b + \gamma_b \gamma_a = 2 \eta_{ab} 1,$
where $\eta$ is diagonal with $p$ pluses and $q$ minuses. We use 
the convention that gamma matrices are unitary so that the ''time'' gammas (first $p$ of them) are 
self-adjoint  while the remaining ones are antiself-adjoint:
$ \gamma_{i}^\dagger = \gamma_i, i=1,\ldots, p$ and
$\gamma_{i}^\dagger = - \gamma_i, i=p+1,\ldots, p+q$.

In even dimensions $p+q = 2d$, we can define: $ \gamma = i^{\frac{p-q}{2}} \gamma_1 \gamma_{2} \cdots \gamma_{p+q}$,
so that $\gamma$ is self-adjoint and $\gamma^2=1$. Now, from the properties of Clifford algebras
we know that there exists a linear unitary operator $B$, such that $B \gamma_i = \epsilon \gamma_i^* B$ and $B B^* = \epsilon'$.
If $B$ is combined with complex conjugation on spinors to give an antilinear operator $J$:
$ J \psi = B \psi^*, $
then we have:
$ J^2 = \epsilon'$ and $JD =  \epsilon DJ, $
where $D$ is the Dirac operator (\cite{bau}, Satz 3.1):
$ D = -\sum_j \eta^{jj} \gamma_j \partial_j.$ 
In the case of even $p+q$ we additionally have:
$ B \gamma = \epsilon'' \gamma B$, where all signs are taken from the Table \ref{table1}.

There exists also another unitary operator $A$ such that
$ A \gamma_i A^\dagger = (-1)^{p+1} \gamma_i^\dagger,$ 
which satisfies:
$ A^2 = (-1)^{\frac{1}{2} p(p-1)}$, $ A \gamma = (-1)^{p} \gamma A$, and $A^* B = \epsilon^p B A. $

The simplest choice for $A$ is $A=\gamma_1...\gamma_p$. Note that the Dirac operator $D$ 
will be $A$ (anti)self-adjoint in the following sense:
\begin{equation} A D A^\dagger = A \sum_j(-\eta^{jj})\gamma_j \partial_j A^\dagger =
(-1)^{p+1} \sum_j(-\eta^{jj})\gamma_j^\dagger \partial_j = (-1)^p D^\dagger.
\end{equation}
To translate the notation to that used in \cite{park} we define
$ \beta = i^{\frac{1}{2} p(p-1)} A$, 
and then we have:
$ \beta= \beta^\dagger$, $\beta^2 = 1$, $\beta \gamma = (-1)^p \gamma \beta$, and
$J \beta = (-1)^{\frac{1}{2} p(p-1)} \epsilon^p \beta J$,
where the last one follows from the observation that
$J\beta=\frac{\alpha^\ast}{\alpha}\epsilon^p\beta J$,
with $\beta=\alpha A$.

The condition for the Dirac operator translates then to:
$ \beta D \beta = (-1)^p D^\dagger$.

As a last remark, we note that the existence of $\beta$ is equivalent to having the Krein product $(\cdot,\cdot)_\beta$ on the Hilbert space, 
where the scalar product and Krein product are related through $(\psi,\phi)=(\psi,\beta\phi)_\beta$.
Then the (essentially) Krein self-adjointness of the operator $T$ is equivalent to the following condition for the Hilbert 
adjoint $T^\dagger$, $T^\dagger=\beta T\beta$. Therefore we see that the operator $i^pD$ is (essentially) Krein self-adjoint, which is 
consistent with \cite{bau}-Satz.3.17, 3.19 and \cite{dun}- Thm. 3.17.  

\subsection{Riemannian triples from pseudo-Riemannian}
Let $(\mathcal{A},\pi,\mathcal{H},D,J,\gamma,\beta)$ be a pseudo-Riemannian spectral 
triple of signature $(p,q)$. Since from the beginning we are working with the 
Hilbert space representation, a passage to the Riemannian spectral triple appears easy. 
Define $D_+=\frac{1}{2}(D+D^\dagger)$ and $D_- = \frac{i}{2}(D-D^\dagger)$. Both $D_\pm$ are
by definition self-adjoint and since $D\gamma= -\gamma D$ and $\gamma$ is self-adjoint then 
also $D^\dagger$ anticommutes with $\gamma$,  therefore ${D_\pm}\gamma=-\gamma {D_\pm}$. 
Using the $\beta$-self-adjointess we see that also $D^\dagger J =\epsilon J D^\dagger$. This
means that $ J D_+ = \epsilon D_+ J$ but (as $J$ is antilinear) $J D_- = - \epsilon D_- J$.
The first order condition also holds for both $D_\pm$, since it holds for $D$ and $D^\dagger$: 
$[D^\dagger,\pi(b)]=(-1)^p\beta[D,\pi(b)]\beta$, and $\beta J\pi(a)J^{-1}\beta=J\pi(a)J^{-1}$ 
for any $a,b\in\mathcal{A}$.
	
As a result we obtain a pair of Riemannian real spectral triples, $(\mathcal{A},\pi,\mathcal{H},D_\pm,J,\gamma)$ which, 
however, differ by $KO$-dimension. Each of them has an additional grading $\beta$, which commutes or 
anticommutes with the Dirac operator: $\beta D_\pm = \pm (-1)^p D_\pm \beta$ and satisfies the conditions 
$\beta^2=1, \hspace{0.1in} \;  \beta^\dagger=\beta$ and $\beta\gamma=(-1)^p\gamma\beta\; , 
\beta J=(-1)^{\frac{p(p-1)}{2}}\epsilon^pJ\beta$.
It is worth noting that in many cases the obtained triples are degenerate in the sense that the kernel of the commutator with each $D_\pm$ is bigger than $\mathbb{C} \subset {\mathcal A}$. 

Yet using both $D_+$ and $D_-$ we can reconstruct a Riemannian spectral triple. Let us define $J_E = J \beta$. Clearly, this is still an antiunitary operator that satisfies:
\begin{equation} J_E^2 = \epsilon' \epsilon^p (-1)^\frac{p(p-1)}{2}, \qquad J_E \gamma = (-1)^p \epsilon'' \gamma J_E.
\end{equation}
Furthermore, we have $ J_E D = (-1)^p \epsilon D^\dagger J_E$ and therefore for both $D_+,D_-$:
\begin{equation} J_E D_\pm = (-1)^p \epsilon D_\pm. 
\end{equation}
As a result, we see that with  the choice $D_E=D_+ + D_-$, $(\mathcal{A},\pi,\mathcal{H},D_E ,J_E,\gamma)$ becomes a Riemannian spectral triple. To match the signs in the Table \ref{table1} for the right $KO$-dimension we might, however, 
need to take (depending on $p$) $J_E' = J_E \gamma$, which would guarantee that 
for the even-dimensional triples we recover the appropriate convention for the signs.  
	
For odd $KO$-dimensions of the pseudo-Riemannian spectral triple we always 
get a Riemannian spectral triple with $J_E$ as a real structure whereas
for even $KO$-dimensions we choose $\widetilde{J_E}=J_E$ as a real structure for $p$ even and $\widetilde{J_E}= J_E'$ for $p$ odd. 

The resulting values of $KO$-dimensions of the Riemannian
spectral triple $(\mathcal{A},\pi,\mathcal{H},D_E ,\widetilde{J_E},\gamma)$, dependent 
on the value of $p \pmod{4}$, are collected in the Table \ref{table2} :  
\begin{center}
\begin{table}[h]
	\caption{$KO$-dimension (mod $8$) for the Riemannian triple obtained from the pseudo-Riemannian triple of signature $(p,q)$.}	
	\label{table2}
		\begin{tabular}{ | c || c | c | c | c | c | c | c | c | }\hline 
			\backslashbox{$p \pmod{4}$}{$p\!-\!q \pmod 8$}
			& $\; 0 \;$ & $\; 1\;$&$\; 2\; $&$\; 3\; $&$\; 4\; $&$\; 5\; $&$\; 6\; $&$\; 7\; $\\ \hline \hline
			$0$ & $0$ & $1$ & $2$ & $3$ & $4$ & $5$ & $6$ & $7$\\ \hline
			$1$ & $2$ & $3$ & $4$ & $5$ & $6$ & $7$ & $0$ & $1$\\ \hline
		   $2$ & $4$ & $5$ & $6$ & $7$ & $0$ & $1$ & $2$ & $3$\\ \hline 
		   $3$ & $6$ & $7$ & $0$ & $1$ & $2$ & $3$ & $4$ & $5$\\ \hline 
		\end{tabular}
	\end{table}
\end{center}
	
Note that comparing the result with Table \ref{table1} we see that it is consistent with passing from the signature $(p,q)$ to $(0,-(p+q))$. 

This procedure of obtaining a Riemannian spectral triple from a pseudo-Riemannian one
could be illustrated easily in many commutative and noncommutative examples, 
including the noncommutative torus. For example, it is easy to demonstrate that 
the above procedure applied to the Lorentzian spectral triple constructed for the noncommutative torus $\mathbb{T}^2_\theta$ in \cite{ps-lor}, Sec. 3, gives the 
usual Riemannian spectral triple over $\mathbb{T}^2_\theta$.

A more interesting application, however, is to the finite spectral triples and, in particular,  the one of the standard model.

\section{Finite spectral triples}\label{fst}

A finite spectral triple is a spectral triple over an algebra, which is a finite direct sum of (complex) matrix algebras. From the 
general consideration (see \cite{ps}) we know that for a finite real spectral triple the algebra $\mathcal{A}$ and the
Hilbert space $\mathcal{H}$ can be decomposed as
\begin{equation}
\mathcal{A}\cong \bigoplus\limits_{i=1}^{N} M_{n_i}(\mathbb{C}), \qquad
\mathcal{H}=\bigoplus\limits_{i,j}\mathcal{H}_{ij},
\end{equation}
with $\mathcal{H}_{ij}=\pi(P_i) J\pi(P_j)J^{-1}\mathcal{H}$, where $P_i\in\mathcal{A}$ is
the identity matrix in the $i$ th entry and zeroes elsewhere. Each of the subspaces $\mathcal{H}_{ij}$
could be written as $\mathcal{H}_{ij}=\mathbb{C}^{n_i}\otimes \mathbb{C}^{r_{ij}}\otimes \mathbb{C}^{n_j},$
where $r_{ij}\in\mathbb{N}$ with the convention that $r_{ij}=0$ means $H_{ij}=0$. The representation $\pi$
is $\pi(a)\xi_{ij}=\left( a_i\otimes 1\otimes 1\right)\xi_{ij}$, where 
$a_i=P_ia=aP_i$ and $\xi_{ij} \in \mathcal{H}_{ij}$. The opposite representation (conjugated by $J$) is
$J \pi(a^*) J^{-1} \xi_{ij}=\xi_{ij}\left(1\otimes 1\otimes a_j^T \right),$ where the latter denotes
the matrix multiplication from the right and $T$ is the matrix transposition.  

We can recall here some of the results of \cite{ps}, which do not depend on the fact that we extend the
triple to be (possibly) pseudo-Riemannian. First, in case $p\!+\!q$ is even, there exists a grading
$\gamma$, which when restricted to $\mathcal{H}_{ij}$ is determined by an internal self-adjoint grading
$\Gamma_{ij}$ on $\mathbb{C}^{r_{ij}}$: $\gamma|_{\mathcal{H}_{ij}}=\gamma_{ij}=1_{n_{i}}\otimes \Gamma_{ij}\otimes 1_{n_j}$.
The antiunitary operator $J$ maps $\mathcal{H}_{ij}$ onto $\mathcal{H}_{ji}$ and therefore $r_{ij}=r_{ji}$.
Observe that $\gamma_{ij}=\epsilon''\gamma_{ji}$ and therefore  $q_{ij}:=r_{ij}\gamma_{ij}$
is a matrix which is symmetric for $KO$-dimensions $0$ and $4$ and antisymmetric for $KO$-dimensions 
$2$ and $6$.

The construction of the Dirac operator follows again the procedure of \cite{ps} and (apart from one condition)
is again independent of the signature. First, we define a linear map $D_{ij,kl}:\mathcal{H}_{kl}\rightarrow \mathcal{H}_{ij}$,
which is $D_{ij,kl}= ( \pi(P_i) J\pi(P_j)J^{-1} ) \, D \, ( \pi(P_k) J\pi(P_l)J^{-1} )$. 

Using the first order condition we conclude (using the same arguments as in \cite{ps}) that the components $D_{ij,kl}$ vanish 
unless $i=k$ or $j=l$. If $i=k$ then $D$ commutes with the representation $\pi$ and if $j=l$ then it commutes with $\pi^\circ$.
Furthermore, we have an analogue of Lemma 7 from \cite{ps}: there exists $\xi=\sum_{i\neq j}P_idP_j$ such that for every 
$a\in\mathcal{A}$ we have $da=[\xi,a]$ and moreover, for the Dirac operator we have $D = \pi(\xi) + \epsilon J\pi(\xi)J^{-1} + \delta$.
Here $\delta$ denotes the part of the Dirac operator (which, in principle can exists) that commutes both with the algebra as well as
with the opposite algebra $J \mathcal{A} J^{-1}$. The (anti)commutation relation with $J$ enforces $D_{ij,kl}=\epsilon J D_{ji,lk}J^{-1}$
We note that observations 5.-7. and Lemmas 8.-11. from \cite{ps} remains unchanged, since the proofs depend only on the properties of 
an algebra $\mathcal{A}$ and the fact that $da=[\xi,a]$. Moreover, the discussion in B.2 of \cite{ps} remains unchanged, 
since it uses only such properties of Dirac operator which do not depend on the signature.

\subsection{The Krein structure for finite spectral triples}

So far we have not considered the existence of $\beta$. Let us check what 
this requirement implies for the rest of the spectral triple. First of all, observe 
that since $\beta$ commutes with the algebra, it is uniquely determined by a family of 
$\beta_{ij}: H_{ij} \to H_{ij}$ such that: $ \beta_{ij} = \beta_{ij}^\dagger, \, \beta_{ij}^2 = 1$. 

The first important statement concerns the existence of $\beta$ for odd $p$, which can be formulated 
in the following way. If at least one subspace $H_{ij}$ such that $r_{ij}>0$ has $\gamma_{ij} = \pm 1$ then 
there exists no pseudo-Riemannian spectral triple on it with $p$ odd. Indeed, in the case of odd $p$, 
we have: $ \beta_{ij} \gamma_{ij}  	= - \gamma_{ij} \beta_{ij}$, so if $\gamma_{ij}$ is proportional 
to the identity matrix we have necessarily $\beta_{ij}=0$ which contradicts $\beta_{ij}^2=1$.

Since $D$ is $\beta$-self-adjoint and commutes with the representation $\pi \;  $ $D_{ij,kl}^\dagger = (-1)^p\beta D_{kl,ij}\beta$,  
for $p=2k$ with $k\in \mathbb{N}$ we have $\beta J=(-1)^k J\beta$ and since for even spectral triples we 
have $p+q\in 2\mathbb{Z}$ then also the $KO$-dimension is even, and therefore $DJ=JD$. 

\subsection{The Riemannian part of finite spectral triples}

The procedure to obtain a Riemannian finite spectral triple from a pseudo-Riemannian one is again 
straightforward, yet the Riemannian spectral triples that are associated with $D_+$ and $D_-$ are 
potentially interesting as they both have an extra symmetry $\beta$. Note that not every 
pseudo-Riemannian is possible, as if $\gamma_{ij}=\pm 1$ for at least one pair $i,j$ then necessarily $p$ needs to
be even as on the subspace $H_{ij}$ we must have $\beta\gamma=\gamma\beta$. 

Nevertheless, bearing in mind that limitation, one can reformulate the problem of constructing a pseudo-Riemannian 
real finite spectral triple as equivalent to the construction of two Riemannian finite-dimensional real 
spectral triples together with a $\mathbb{Z}_2$-grading $\beta$ that satisfies certain commutation 
relations with $\gamma$ and $J$ and Dirac operators that commute or anticommute with $\beta$.

A simple example of this, could be illustrated by a finite spectral triple over 
an algebra with two summands. For simplicity we could take them both to be 
$\mathbb{C}$, so that $\mathcal{A} = \mathbb{C} \oplus \mathbb{C}$ 
and $H_{ij} = \mathbb{C}$  for $i,j=1,2$. This is the basic setup leading to the 
real spectral triple for an algebra of functions over two points. We should 
remark here that the Hilbert space is $\mathbb{C}^4$ and we can easily write
the representation and all operations using matrices in $M_4(\mathbb{C})$. 
Identifying $\mathbb{C}^4$ as $H_{11} \oplus H_{21} \oplus H_{12} \oplus H_{22}$
we have $\pi$, $J$, $\gamma$:
\begin{equation}
\pi(z \oplus w) =
\mleft(
\begin{array}{cccc}
\; z & \; 0 & \; 0 & \; 0 \\
\; 0 & z & \; 0 & \; 0 \\
\; 0 & \; 0 & w & \; 0 \\
\; 0 & \; 0 & \; 0 & \; w\\
\end{array}
\mright), \quad
\gamma = \mleft(
\begin{array}{cccc}
\; 1 & \; 0 & \; 0 & \; 0 \\
\; 0 & -1 & \; 0 & \; 0 \\
\; 0 & \; 0 & -1 & \; 0 \\
\; 0 & \; 0 & \; 0 & \; 1\\
\end{array}
\mright), \quad
J = \mleft(
\begin{array}{cccc}
\; 1 & \; 0 & \; 0 & \; 0 \\
\; 0 & \; 0 & \; 1 & \; 0 \\
\; 0 & \; 1 & \; 0 & \; 0 \\
\; 0 & \; 0 & \; 0 & \; 1
\end{array}
\mright) \circ \ast.
\end{equation}
Now, we can easily identify a nontrivial additional symmetry ($\mathbb{Z}_2$-grading) $\beta$ and construct a Dirac
operator $D_+$, which is real, satisfies first-order condition and commutes with $\beta$:
\begin{equation}
\beta = \mleft(
\begin{array}{cccc}
\; 1 & \; 0 & \; 0 & \; 0 \\
\; 0 & 1 & \; 0 & \; 0 \\
\; 0 & \; 0 & 1 & \; 0 \\
\; 0 & \; 0 & \; 0 & -1 \\
\end{array}
\mright), \quad
D_+ = \mleft(
\begin{array}{cccc}
\;  0 & \; d & \; d^* & \; 0 \\
\; d^* & \; 0 & \; 0 & \; 0 \\
\; d & \; 0 & \; 0 & \; 0 \\
\; 0 & \; 0 & \; 0 & \; 0
\end{array}
\mright).
\end{equation}

This data gives us a Riemannian real spectral triple of $KO$-dimension $0$ over the algebra of functions
on two points, which has an additional symmetry. In contrast to the full Riemannian spectral triple, where the 
Dirac operator has two arbitrary complex entries (see \cite{ds} for details) the restriction due to the 
$\beta$-symmetry gives only one free parameter into the family of possible Dirac operators.

It is an easy exercise, which we omit, to construct a second Riemannian spectral triple, with a Dirac operator
$D_-$, which, in addition would satisfy $D_- \beta = - \beta D_-$. Both these triples could be seen as Riemannian parts 
of a pseudo-Riemannian spectral triple with signature $(4,4)$ or $(0,0)$. Again, it is easy to check that the
full pseudo-Riemannian real spectral triple over the algebra of function on two points would have the full
Dirac operator $D$:
\begin{equation}
D = \mleft(
\begin{array}{cccc}
0 & d & d^* & 0 \\
d^* & 0 & 0 & c \\
d & 0 & 0 & c^* \\
0 & -c^* & -c & 0
\end{array}
\mright),
\end{equation}
where $c,d$ are arbitrary complex numbers.

We shall explore this effect more for the spectral triple of the standard model.

\section{The Standard Model}
\label{IV}
In this section we shall discuss the finite spectral triple for standard model, using the conventions consistent with \cite{wvs,das}. As an algebra we take
\begin{equation}
 A_F=\mathbb{C}\oplus \mathbb{H}\oplus M_3(\mathbb{C}),
 \end{equation}
represented on the Hilbert space
\begin{equation}
H_F=\left(H_l \oplus H_q \right)\oplus\left(H_{\bar{l}} \oplus H_{\bar{q}} \right),
\label{HF}
\end{equation}
where the basis for the leptonic space $H_l$ is ordered as $\{\nu_R,e_R,(\nu_L,e_L)\}$ and the quark space $H_q$ have a basis $\{u_R,d_R,(u_L,d_L)\}$ (in each color). The generic algebra element,
$(\lambda \oplus h \oplus m)$ from $\mathbb{C}\oplus \mathbb{H}\oplus M_3(\mathbb{C})$,
is represented on $H_l$  and $H_q$ (for each color) as
$ \pi(\lambda,h,m) = \lambda \oplus \bar{\lambda}  \oplus  h, $ whereas the
representation on $H_{\bar{l}}$ is by $\lambda$ and on $H_{\bar{q}}$ by $1_4\otimes m$, where we used the fact that we have $3$ colours.
In the case when we consider $N$ generations the Hilbert space is respectively
enlarged by tensoring it with $\mathbb{C}^N$ whereas the representation
is extended diagonally.

The physical Dirac operator in this description of the standard model has a form
\begin{equation}
D_F=\begin{pmatrix}
S & T^\dagger \\
T & \bar{S}
\end{pmatrix}, \qquad \qquad
S=\begin{pmatrix}
S_l & \\
& S_q \otimes 1_3
\end{pmatrix},
\label{DiF1}
\end{equation}

with
\begin{equation}
S_l=
\renewcommand\arraystretch{1.3}
\mleft[
\begin{array}{c|c|c|c}
   &&Y_\nu^\dagger& \\
  \hline
  & & &Y_e^\dagger\\ 
  \hline
  Y_\nu&&& \\
  \hline
  &Y_e&&
\end{array}
\mright], 
\hspace{10pt}
S_q=
\renewcommand\arraystretch{1.3}
\mleft[
\begin{array}{c|c|c|c}
   &&Y_u^\dagger& \\
  \hline
  & & &Y_d^\dagger\\ 
  \hline
  Y_u&&& \\
  \hline
  &Y_d&&
\end{array}
\mright],
\label{DiF2}
\end{equation}
where $Y_\nu,Y_e,Y_u,Y_d$ are Yukawa mass matrices. 
The operator $T$ is given by $T \nu_R=Y_R\bar{\nu_R},$ for a certain 
symmetric Majorana mass matrix $Y_R \in M_N(\mathbb{C})$ and $T$ gives 
zero on other fermions. It is well known that $(A_F,H_F,D_F)$ together with $\gamma_F$ 
acting as $1$ on right-handed and as $-1$ on left-handed particles, and $J_F$ conjugation 
composed with exchanging particles with antiparticles, form a spectral triple.

Note that this spectral triple is nonorientable \cite{ste}. The reason is that we 
have included sterile right-handed neutrinos and any operator of the form 
$\sum_k \pi(\lambda_k,...)J_F\pi(\mu_k,...)J_F^{-1}$ 
acting on $\nu_R$ and $\bar{\nu}_R$ reproduces them with the same value $\sum_k\lambda_k\bar{\mu}_k$, 
but $\gamma_F\nu_R=\nu_R$ and $\gamma_F\bar{\nu}_R=-\bar{\nu}_R$. 

There is no such problem for right electron since the representation on electron differs 
by complex conjugation.  Orientability can be restored if one assumes that there are 
only left-handed neutrinos\footnote{Usually these are in the literature identified with Majorana neutrinos, though the finite spectral triple of the standard model does not 
involve spinorial degrees of freedom.}, then one can easily find a Hochschild cycle 
that provides the orientability, however, with the expense that neutrino masses cannot 
be of the same origin as for other leptons and quarks but arising from an effective 
mass terms such as Weinberg effective Lagrangian \cite{Wei}. It is also well known that 
the above Dirac operator is not unique (see \cite{PSS,das}) within the model-building
scheme of noncommutative geometry. Even the introduction of more constraints, like the second-order condition \cite{das} or Hodge-duality \cite{d} does not allow to exclude the terms, which would introduce the couplings between lepton and quarks and lead to the leptoquark fields \cite{PSS}.

\subsection{The pseudo-Riemannian shadow for the standard model}\label{sec4}

The solution to the problem lies in the introduction of an additional symmetry, in terms of the 
${\mathbb{Z}}_2$-grading, which distinguishes between lepton and quarks. This symmetry, which could be taken
as a $0$-cycle:
\begin{equation}
\beta = \pi(1,1,-1) J_F \pi(1,1,-1) J_F^{-1}, \label{betaF}
\end{equation}
has had various interpretations and originally was linked to K-theoretic origins \cite{c,kraj}. However, in the light of the 
discussion of the pseudo-Riemannian spectral triples we propose, another explanation that uses the 
pseudo-Riemannian construction.

Observe that $\beta$ (\ref{betaF}) satisfies all commutation relations with $\gamma_F$ and $J_F$ that are 
consistent with $p = 4k$. Since the total $KO$-dimension is unchanged when passing to the Riemannian restriction 
with $D_+$ then $q=4k+2 \pmod{8}$. For $p=0$ (i.e. $k=0$) $q$ has to be equal $2 \pmod{8}$, hence the simplest 
choice for the (finite part of) standard model is the pseudo-Riemannian spectral triple of signature $(0,2)$.

Therefore $(A_F,H_F,D_F,\gamma_F,J_F, \beta)$ could be seen as a Riemannian restriction 
of a real even pseudo-Riemannian spectral triple of signature $(0,2)$ [note that this choice is 
not unique and it is also possible to chose in a consistent way e.g. the signature $(4,6)$]. 

\subsection{Classification of pseudo-Riemannian shadows for the standard model.}

The example in the previous section demonstrated that the additional 
symmetry can be interpreted as a $0$-cycle being the shadow
of the pseudo-Riemannian structure. In this section we shall look for 
more general structures of this type, aiming to step towards their 
classification and study the physical consequences.

The approach we take here assumes as a starting point the general structure of the 
finite spectral triple for the standard model as discussed in \cite{das}, however, 
without further restrictions on the Dirac operator than order one condition. Then we 
shall look for all possible operators $\beta$, which are $0$-cycles and commute with
the Dirac operator. This shall lead to constraints on the Dirac operator, which we
shall then compare with the condition of Hodge duality discussed in \cite{das}. 

Let us briefly recall the details of the standard model spectral triple. We take as a Hilbert space $H_F = F\oplus F^\ast$ with vectors from $H_F$ can be represented as a pair
of matrices, $v,w\in M_4(\mathbb{C})$, 
\begin{equation}
\begin{bmatrix}
v\\ w
\end{bmatrix} \in H_F, \quad
v=
\renewcommand\arraystretch{1.3}
\mleft[
\begin{array}{cccc}
\nu_R&u_R^1&u_R^2&u_R^3 \\
e_R&d_R^1&d_R^2&d_R^3 \\
\nu_L&u_L^1&u_L^2&u_L^3 \\
e_L&d_L^1&d_L^2&d_L^3 \\
\end{array}
\mright], \quad
w=
\renewcommand\arraystretch{1.3}
\mleft[
\begin{array}{cccc}
\overline{\nu_R} & \overline{e_R} & \overline{\nu_L} & \overline{e_L} \\ 
\overline{u_R^1} & \overline{d_R^1} & \overline{u_L^1} & \overline{d_L^1} \\
\overline{u_R^2} & \overline{d_R^2} & \overline{u_L^2} & \overline{d_L^2} \\
\overline{u_R^3} & \overline{d_R^3} & \overline{u_L^3} & \overline{d_L^3} 
\end{array}
\mright].
\end{equation}
This presentation is more convenient to describe all possible Dirac operators and symmetries arising from pseudo-Riemannian structures. It is easy to identify $H_l$
and $H_q$ from (\ref{HF}) as the first column and, respectively, three last columns
of the $F$ matrix. The bonus, however, is in the identification of the algebra of
all possible linear transformations of $H_F$, as it could be described as elements
of the $M_4(\mathbb{C})\otimes M_2(\mathbb{C})\otimes M_4(\mathbb{C})$ algebra.
A simple tensor from this algebra $m_L \otimes m \otimes m_R$ acts on the vector
composed from $v,w$ in the following way: $m_L$ acts by left matrix multiplication on
$v,w$;  $m_R$ acts by right matrix multiplication by its transpose (on $v,w$) while 
$m$ acts a linear transformation on the pair $[v,w]$.
If we denote by $e_{ij}$ a matrix with the $1$ in position $(i,j)$ and zero everywhere else,
and use $1_k$ to denote the identity matrix in $M_k(\mathbb{C})$
then we can conveniently write the real structure and the grading $\gamma$ as
\begin{equation}
J\begin{bmatrix}
v\\w
\end{bmatrix}=
\begin{bmatrix}
w^\ast\\ v^\ast
\end{bmatrix}, \qquad 
\gamma=\begin{bmatrix}
1_2 &\\ & -1_2
\end{bmatrix}\otimes e_{11} \otimes 1_4 + 1_4 \otimes e_{22}\otimes \begin{bmatrix}
-1_2& \\ &1_2
\end{bmatrix},
\end{equation}
whereas the elements of the algebra $A=\mathbb{C}\oplus\mathbb{H}\oplus M_3(\mathbb{C})$ are represented on $H_F$ by
\begin{equation}
\pi(\lambda,q,m) =
\sbox0{$\begin{matrix}\lambda & \\&\overline{\lambda}\end{matrix}$}
\left[
\begin{array}{c|c}
\usebox{0}&\makebox[\wd0]{\large $0$}\\
\hline
\vphantom{\usebox{0}}\makebox[\wd0]{\large $0$}&\makebox[\wd0]{\large $q$}
\end{array}
\right]\otimes e_{11}\otimes 1_4 + 
\sbox1{$\begin{matrix}\lambda \end{matrix}$}
\left[
\begin{array}{c|c}
\usebox{1}&\makebox[\wd0]{\large $0$}\\
\hline
\vphantom{\usebox{0}}\makebox[\wd0]{\large $0$}&\makebox[\wd0]{\large $m$}
\end{array}
\right]\otimes e_{22}\otimes 1_4,
\end{equation}
where $\lambda\in \mathbb{C}, q\in\mathbb{H}$ and $m\in M_3(\mathbb{C})$.

The most general Dirac operator is of the form $D=D_0+D_1+D_R,$
where $D_1=JD_0J^{-1}$ and $D_R$ is $J$ invariant. We consider a spectral 
triple of $KO$-dimension $6$, with a self-adjoint Dirac operator, but such that commutes with a suitable $\beta$ that represents the shadow of a pseudo-Riemannian structure. We limit our considerations to the case
of a Dirac operator that satisfies an order-one condition. We have
\begin{equation}
\begin{aligned}
D_0=& \begin{bmatrix}
&M\\ M^\dagger&
\end{bmatrix} \otimes e_{11}\otimes e_{11} + \begin{bmatrix}
&N\\ N^\dagger&
\end{bmatrix} \otimes e_{11}\otimes (1- e_{11})+ \\ & + 
\begin{bmatrix}
A&B\\0&0
\end{bmatrix} \otimes e_{12}\otimes e_{11} + \begin{bmatrix}
A^\dagger&0\\
B^\dagger&0
\end{bmatrix}\otimes e_{21}\otimes e_{11},
\end{aligned}
\end{equation}
where $M,N, A, B$ are all in $M_2(\mathbb{C})$. Observe that the assumed form
of $D_0$ includes already $D_R$:
\begin{equation}
D_R=e_{11}\otimes \left( A_{11} e_{21} + A_{11}^*  e_{12} \right)\otimes e_{11}.
\end{equation}
The presentation of the Dirac operator in \ref{DiF1}--\ref{DiF2} corresponds
to $M=S_l$, $N=S_q$ and $A=T$ (here to have the matrix equality $T$ needs to
be restricted to the subspace of right-handed leptons), $B=0$. 

In the next step of our search for all possible shadows of pseudo-Riemannian
structures we look for a $\beta$ that is a $0$-cycle, i.e. a sum of elements 
of the form
\begin{equation}
\beta = \pi(\lambda_1,q_1,m_1) J \pi(\lambda_2,q_2,m_2) J^{-1}, 
\end{equation}
with $\lambda_1,\lambda_2\in \mathbb{C},\; q_1,q_2\in \mathbb{H},\; m_1,m_2\in M_3(\mathbb{C})$. As we impose the condition that $\beta^2=1$, $J\beta = \beta J$ 
and $\beta$ commutes with the algebra, we obtain that all elements $\lambda_1,\lambda_2$ as well as the matrices $q_1, q_2$ and $m_1,m_2$ must 
be $\pm 1$ and pairwise equal to each other (that is $\lambda_1\!=\!\lambda_2$, $q_1\!=\!q_2$, $m_1\!=\!m_2$). Therefore, up to a trivial rescaling (by $-1$), we have three possibilities. The one discussed in the preceding subsection is $\lambda_1=1$, 
$q_1 \!=\! 1_2$ and $m_1\!=\!- 1_3$. 

Let us discuss all three cases. First, if $\beta=\pi(1,-1,1) J\pi(1,-1,1) J^{-1}$ then 
the restrictions for the Dirac operator to commute with $\beta$ are only $M=N=0$, with
no restriction for $A,B$.  

The second case, with  $\beta=\pi(-1,1,1) J\pi(-1,1,1) J^{-1}$ imposes similarly $M,N=0$,
but then additionally $B=0$ and  $A$ that has to satisfy $A = A \cdot \hbox{diag}(1,-1)$.  

The last case, with $\beta=\pi(1,1,-1) J\pi(1,1,-1) J^{-1}$, which was discussed in the
preceding sections enforces $B=0$ and $A = A \cdot \hbox{diag}(1,-1)$ while putting
no constraints on $M$ and $N$, consistent with the discussion at the beginning of
Secs. \ref{IV} and \ref{sec4}.  Note that the fact that $A$ does not vanish is consistent
with the existence of terms that involve only the sterile neutrino and thus compatible
with physical data.

It is worth noting that both only the last situation allows for a physical Dirac operator
(with Majorana mass terms for the neutrinos) and, moreover, in the view of the results of
\cite{das} it is the only case that satisfies the Hodge duality.

Therefore, as a consequence we see that the only possible $0$-cycle for a real spectral
triple over the standard model that can be interpreted shadow of a pseudo-Riemannian structure, which additionally allows Hodge duality is the one with $\beta = \pi(1,1,-1) J\pi(1,1,-1) J^{-1}$, resulting in the symmetry that physically is interpreted as lepton number conservation. 

\section{Conclusions}

As the finite-dimensional spectral triple of the standard model shares the property of degeneracy with the Riemannian parts of the pseudo-Riemannian 
triples, i.e. the $SU(3)$ symmetry remains unbroken due to the apparent preserved symmetry between leptons and quarks, we conjecture that this is the  
genuine origin of that feature. Of course, to consider the full standard model as an almost commutative geometry we need to construct the product of 
the standard triple over a manifold $M$ with the above-defined triple. This could be in the Lorentzian setup only on an algebraic level, as the analytic
tools leading to the spectral action and description of Yang-Mills and Higgs terms in the action can be carried out only in the Euclidean framework.

Moreover, the product geometry leads to the problem of fermion doubling \cite{ferm}, which has been discussed in both Euclidean and Lorentzian setup. 
However, the interpretation of the finite part as a Riemannian part of some possibly bigger, pseudo-Riemannian finite geometry can shed a new light
on the issue, which we shall tackle in the next work. Leaving these problems aside we can, however, offer a very natural interpretation of the 
preserved symmetry between the leptons and the quarks as originating from the much deeper, pseudo-Riemannian geometrical structure of the 
standard model. Together with the recently uncovered property of Hodge duality \cite{das} this allows to eliminate all Dirac operators that would
mix leptons and quarks and, in consequence lead to leptoquarks and breaking of the $SU(3)$ symmetry. It offers also an intriguing possibly to
look for the full pseudo-Riemannian triple, which can be a next step to reach beyond the standard model.

\section*{Acknowledgements} The authors acknowledge support by NCN Grant No. OPUS 2016/21/B/ST1/02438.



\begin{thebibliography}{20}
\bibitem{con} A.~Connes and M.~Marcolli, {\it Noncommutative Geometry, Quantum Fields and Motives} (Colloquium Publications, 2008) Vol. {\bf 55}.
\bibitem{madore} J.~Madore, {\it Classical gravity on fuzzy space-time}, Nucl.Phys.Proc.Suppl. \textbf{56B},  183 (1997).
\bibitem{grav}  A.H.~Chamseddine, A.~Connes and M.~Marcolli, {\it Gravity and the standard model with neutrino mixing}, Adv. Theor. Math. Phys. \textbf{11}, 991 (2007).
\bibitem{wvs} W.D.~van Suijlekom, {\it Noncommutative Geometry and Particle Physics} (Springer, New York, 2015).
\bibitem{ps} M.~Paschke and A.~Sitarz, {\it Discrete spectral triples and their symmetries}, J. Math. Phys. {\bf 39}, 6191 (1998).
\bibitem{kraj} T.~Krajewski, {\it Classification of finite spectral triples}, J. Geom. Phys. \textbf{28}, 1 (1998).
\bibitem{das} L.~Dąbrowski, F.~D'Andrea and A.~Sitarz, {\it The standard model in noncommutative geometry: Fundamental fermions as internal forms}, Lett. Math. Phys. (2017). 
\bibitem{ps-lor} M.~Paschke and A.~Sitarz, {\it Equivariant Lorentzian spectral triples}, arXiv:math-ph/0611029.
\bibitem{barret} J.~Barret , {\it Lorentzian version of the noncommutative geometry of the standard model of particle physics}, J. Math. Phys. {\bf 48}, 012303 (2007).
\bibitem{franco} M.~Eckstein and N.~Franco, {\it Noncommutative geometry, Lorentzian structures and causality}, arXiv:1409.1480v1.
\bibitem{lizzi1} F.~Lizzi, G.~Mangano, G.~Miele and G.~Sparano, {\it Fermion Hilbert space and fermion doubling in the noncommutative geometry approach to gauge theories}, Phys. Rev. D {\bf 55} 6357 (1997).
\bibitem{doubl} J.M.~Gracia-Bond\'ia, B.~Iochum and T.~Schucker, {\it The standard model in noncommutative geometry and fermion doubling}, Phys. Lett. B {\bf 416} 123 (1998).
\bibitem{boyle} L.~Boyle and S.~Farnsworth, {\it A new algebraic structure in the standard model of particle physics}, arXiv:1604.00847.
\bibitem{bizi} F.~Besnard, N.~Bizi and CH.~Brouder., {\it The standard model as an extension of the noncommutative algebra of forms}, arXiv:1504.03890.
\bibitem{d} L.~Dąbrowski, {\it On noncommutative geometry of the Standard Model: fermion multiplet as internal forms}, arXiv:1711.06492.
\bibitem{PSS} M.~Paschke, F.~Scheck and A.~Sitarz, {\it Can (noncommutative) geometry accommodate leptoquarks?}, Phys. Rev. D {\bf 59}, 035003 (1999).
\bibitem{c} A.~Connes, {\it Gravity coupled with matter and the foundation of noncommutative geometry}, Comm.Math.Phys, 155,109 (1996).
\bibitem{co} A.~Connes, {\it Non-commutative geometry and reality}, J.Math.Phys. {\bf 36} 11 (1995).
\bibitem{pd} C.~Patrignani {\it et al.} (Particle Data Group), {\em The review of particle physics (2017)}  Chin. Phys. C {\bf 40}, 100001 (2016) [and 2017 update].
\bibitem{bau} H.~Baum, {\it Spin-Strukturen  und  Dirac-Operatoren \"uber  pseudo-Riemannschen  Mannigfaltigkeiten}, Teubner-Texte zur Mathematik (Teubner-Verlag, Leipzig, 1981) vol. 41.
\bibitem{park}  J.-H.~Park, {\it Lecture note on Clifford algebra}, Lecture at Modave Summer School in Mathematical Physics, Belgium, 06., URL. 124, 129 (2005).
\bibitem{dun} K.L.~van den Dungen, {\it Lorentzian geometry and physics in Kasparov's theory}, PhD thesis Ph.D. thesis in Mathematical Sciences, The Australian National University, 2015. 
\bibitem{ds} L.~Dąbrowski and A.~Sitarz, {\it Twisted reality condition for spectral triple on two points,} Prod. Sci. CORFU2015 ({\bf 2015}) 093 .
\bibitem{ste}  C.~Stephan, {\it Almost-commutative geometry, massive neutrinos and the orientability axiom in KO-dimension 6} arXiv:hep-th/0610097.
\bibitem{Wei} S.~Weinberg, {\it Baryon- and Lepton-Nonconserving Processes}, Phys. Rev. Lett. {\bf 43}, 1566, (1979).
\bibitem{ferm} F.~D'Andrea, M.~Kurkov and F.~Lizzi, {\it Wick rotation and fermion doubling in noncommutative geometry}, Phys. Rev. D {\bf 94}, 025030 (2016).


\end{thebibliography}
\end{document}